\documentclass[structabstract,bibyear]{aa}
\usepackage{graphicx}
\usepackage{txfonts}

\usepackage[authoryear,round]{natbib}

\usepackage{color}

\begin{document}
   \title{Tensile \& shear strength of porous dust agglomerates}


   \author{A. Seizinger,
          \inst{1}
          R. Speith,
          \inst{2}
          \and
          W. Kley\inst{1}
          }

   \institute{Institut f\"ur Astronomie and Astrophysik, Eberhard Karls Universit\"at T\"ubingen,\\
              Auf der Morgenstelle 10, D-72076 T\"ubingen, Germany\\
              \email{alexs@tat.physik.uni-tuebingen.de}
              \and
              Physikalisches Institut, Eberhard Karls Universit\"at T\"ubingen,\\
              Auf der Morgenstelle 14, D-72076 T\"ubingen, Germany\\
             }

   \date{Received 10.06.2013; accepted 23.08.2013}

 
  \abstract
   {Within the sequential accretion scenario of planet formation, planets are build up through a sequence sticking collisions.
  The outcome of collisions between porous dust aggregates is very important for the growth from very small dust particles to planetesimals.
  In this work we determine the necessary material properties of dust aggregates as a function the porosity.}
   {Continuum models such as SPH that are capable of simulating collisions of macroscopic dust aggregates require a set of material parameters. Some of them such as the tensile and shear strength are difficult to obtain from laboratory experiments. The aim of this work is to determine these parameters from ab-initio molecular dynamics simulations.}
   {We simulate the behavior of porous dust aggregates using a detailed micro-physical model of the interaction of spherical grains that includes adhesion forces, rolling, twisting, and sliding. Using different methods of preparing the samples we study the strength behavior of our samples with varying
 porosity and coordination number of the material.}
	{For the tensile strength, we can reproduce data from laboratory experiments very well. For the shear strength, there are no experimental data available. The results from our simulations differ significantly from previous theoretical models, which indicates that the latter might not be sufficient to describe porous dust aggregates.}
   {We have provided functional behavior of tensile and shear strength of porous dust aggregates as a function of the porosity that can be directly
   applied in continuum simulations of these objects in planet formation scenarios.}


	\authorrunning{Seizinger et al.}
	\maketitle

\section{Introduction}

The formation of planetesimals, km sized objects that are massive enough for gravity to come into play, constitutes a key step of the core accretion scenario for planet formation proposed by \citet{1996Icar..124...62P}. However, the earlier growth from mm to km sized bodies is not fully understood yet. Understanding the interplay between porosity, impact velocity, and the size and structure of colliding aggregates in the meter-size regime is crucial to unravel the process of planetesimal formation. Obviously, this size regime renders laboratory experiments impossible. Thus, for years to come astrophysicists will have to rely on computer simulations to obtain the necessary insight into this complex process.

Molecular dynamics simulations featuring detailed micro-mechanical interactions have been employed to study collisions of sub-mm sized dust and ice aggregates \citep[e.g.][]{1997ApJ...480..647D, 2007ApJ...661..320W, 2012ApJ...752..151R}. Owing to the high computational demand a different approach is necessary for the mm to meter size regime. In this regime, SPH simulations have often been utilized to model pre-planetesimal collisions \citep[e.g.][]{2004Icar..167..431S, 2007A&A...470..733S, 2011A&A...536A.104G}. Smoothed particle hydrodynamics (SPH) constitutes a continuum approach that is capable to simulate the collisional behavior of macroscopic aggregates including physical processes such as compaction or fragmentation. Being a continuum approach, SPH requires various material parameters such as the compressive, tensile, and shear strength, and hence a proper calibration is necessary \citep{2010A&A...513A..58G}. Typically, the calibration process is based on comparison with results from laboratory experiments \citep{2009ApJ...701..130G}. However, not all material parameters have been obtained in this way. E.g.\,the shear strength is based only on theoretical models and estimations so far.

So far, only few laboratory experiments have been performed to investigate the mechanical properties of porous dust aggregates. \citet{PhysRevLett.93.115503} measured the tensile strength of highly porous dust aggregates generated by random ballistic deposition. As they used monodisperse, spherical silica grains their experiments are comparable to our simulations. It has been shown both theoretically \citep{2009A&A...504..625B} and experimentally \citep{2006ApJ...652.1768B} that the mechanical properties depend on the shape and size distribution of the grains. Recently, \citet{2012A&A...544A.138M} presented results from various experiments on the mechanical properties of irregularly shaped quartz aggregates.

Determining material parameters directly from molecular dynamics simulations of porous dust aggregates constitutes a tempting alternative. \citet{2008A&A...484..859P} presented the first attempt to obtain the compressive strength from ab-initio simulations. A few years later, \citet{2012A&A...541A..59S} studied the compressive strength in greater detail and especially revealed the differences between static and dynamic compaction processes. 

The aim of the present work is to extend this approach to determine the tensile and shear strength of porous aggregates. Together with the compressive strength we then can provide continuum simulations (such as SPH) with a complete parameter set describing the transition from elastic to plastic deformation.

\section{Interaction model}

\label{sec:interaction_model}

We simulate the behavior of porous dust aggregates with a molecular dynamics approach. In this work we use cuboidal shaped aggregates with an edge length of $30$ to $60\,\mathrm{\mu m}$. Depending on its volume filling factor such an aggregate consists of up to $6\cdot10^4$ micron sized spherical grains (monomers). 

The interaction between individual monomers is based on the work of \citet{1997ApJ...480..647D}. Based on earlier theoretical work by \citet{1971RSPSA.324..301J,1995PMagA..72..783D, 1996PMagA..73.1279D}, they developed a detailed micro-mechanical model of the interaction of two microscopic spherical grains. When two monomers touch each other, surface forces allow for the creation of an adhesive contact. Upon deformation of these contacts caused by the relative motion of the monomers kinetic energy is dissipated. This approach is favorable for our purpose because the process of internal restructuring of the aggregate is modeled far more realistically compared to simpler hard-sphere models.

Later, \citet{2007ApJ...661..320W} derived almost the same interaction laws from corresponding potentials. \citet{2012A&A...541A..59S} calibrated the interaction model by comparison with laboratory experiments on the compression of porous dust aggregates \citep{2009ApJ...701..130G}. They observed that the original model of \citet{1997ApJ...480..647D} was too soft. In order to increase the strength of the aggregates, \citet{2012A&A...541A..59S} introduced the rolling and sliding modifiers $m_\mathrm{r}$ and $m_\mathrm{s}$ that modify the strength of the corresponding type of interaction. By increasing the rolling interaction by a factor of $8$ and the sliding interaction by a factor of $2.5$ they achieved excellent agreement between numerical simulations and laboratory results.

In this work we use the modified interaction model presented by \citet{2012A&A...541A..59S} with $m_{\mathrm{r}} = 8$ and $m_{\mathrm{s}} = 2.5$ unless stated otherwise. The material parameters are listed in Tab.\,\ref{tab:material_parameters}.

\begin{table}
 \caption[]{Material Parameters.}
 \label{tab:material_parameters} 
 \centering
 \renewcommand\arraystretch{1.2}
 \begin{tabular}{ll}
   \hline
   \noalign{\smallskip}
   Physical property & Silicate\\
   \noalign{\smallskip}
   \hline
   \noalign{\smallskip}
   Particle Radius $r$ (in $\mathrm{\mu m}$) & $0.6$\\
   Density $\rho$ (in g\, cm$^{-3}$) & $2.65$\\
   Surface Energy $\gamma$ (in mJ\, m$^{-2}$) & $20$\\
   Young's Modulus $E$ (in GPa)         & $54$\\
   Poisson Number $\nu$               & $0.17$\\
   Critical Rolling Length $\xi_\mathrm{crit}$ (in nm) & $2$\\
   \noalign{\smallskip}
   \hline
 \end{tabular}
\end{table}

\section{Sample generation}

\begin{figure}
\resizebox{\hsize}{!}{\includegraphics{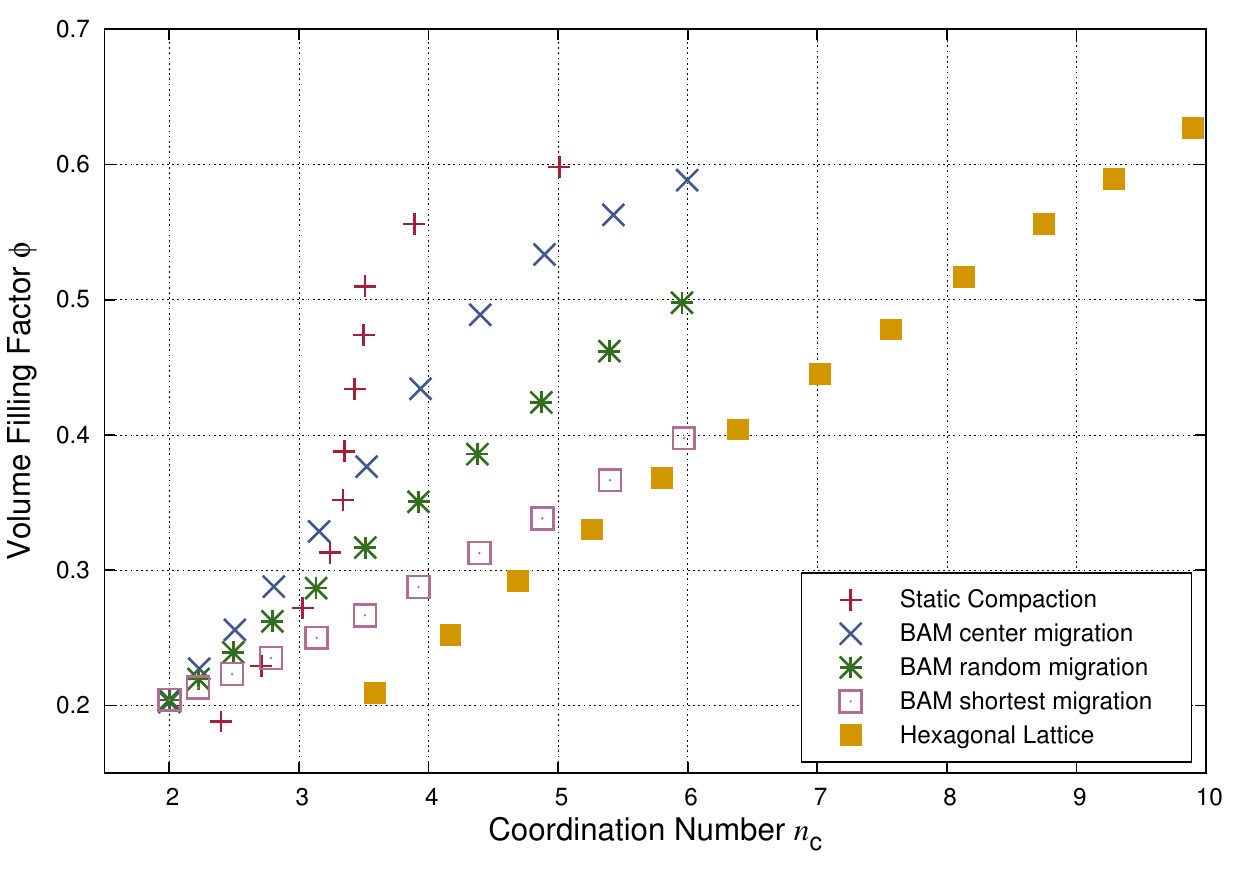}}
\caption{The relation between the volume filling factor $\phi$ and the average coordination number $n_\mathrm{c}$. Figure taken from \citet{2013A&A...551A..65S}.}
\label{fig:ff_nc}
\end{figure}

In this Section we briefly summarize our sample generation methods. Here we use cuboidal shaped dust aggregates (also referred to as dust cakes) of different porosities. In principle, we may employ different methods to generate these samples. As shown in in Fig.\,\ref{fig:ff_nc} the relation between the volume filling factor and the average coordination number $n_\mathrm{c}$ depends on the selected generation method \citep{2013A&A...551A..65S}. 

The volume filling factor $\phi$ is given by 
\begin{eqnarray}\phi = \frac{N V_\mathrm{p}}{V},\end{eqnarray}
where $N$ denotes the number of monomers, $V_\mathrm{p}$ is the volume of an individual  monomer, and $V$ is the total volume occupied by the sample. The coordination number, $n_\mathrm{c}$, denotes the number of contacts a monomer has established with its neighbors. The mean $n_\mathrm{c}$ of the sample is calculated by averaging the number of contacts of each monomer.

\subsection{Static compaction}
Static compaction constitutes a method that very closely resembles the generation process of dust cakes in laboratory experiments. Initially, a dust cake is built by random ballistic deposition (RBD) where single monomers are successively dropped onto the existing sample. They come to rest right at the spot where they hit the existing sample. This growth process results in fractal, highly porous aggregates with a volume filling factor of $\phi = 0.15$ \citep{Watson:1997}.

In the second step, the RBD cake is put into a box and compacted until the desired filling factor is reached. However, the compaction must occur slowly enough in order to avoid inhomogeneities \citep{2012A&A...541A..59S}. For the material/monomer size used in this work a typical speed of the compacting wall is $1\,\mathrm{cm\,s^{-1}}$. The simulation time needed to compact an aggregate is proportional to the desired size and filling factor. Since the number of monomers increases with the size and compactness of the final dust cake the computational effort per integration step rises as well. Thus, generating samples by static compaction can become a computationally expensive, time consuming procedure.

Elastic loading of dust cakes compressed to filling factors above $\approx 0.45$ constitutes another setback of this method \citep{2013A&A...551A..65S}. Because of the elastic energy stored in the contacts between monomers the aggregate will start to expand once the confining walls of the compaction box are removed. This effect will alter the results of a measurement of the tensile or shear strength significantly.

Up to a filling factor of $\approx 0.58$ dust cakes may be stabilized in the following way: After slightly disturbing the positions of monomers the aggregate is kept in a box until the energy induced by the disturbance is damped away \citep{2013A&A...551A..65S}.

\subsection{Ballistic aggregation and migration}
The generation procedure of ballistic aggregation and migration (BAM) has been proposed by \citet{2008ApJ...689..260S}. A larger aggregate is generated by successively shooting in single monomers from random directions onto the existing aggregate.

In \citet{2013A&A...551A..65S} we employed three different methods to select the final position of a monomer hitting the aggregate:
\begin{enumerate}
 \item Select the position closest to the spot, where the monomer impacts on the aggregate (referred to as ``shortest migration'').
 \item Select the position randomly from all available possibilities (referred to as ``random migration'').
 \item Select the position which is closest to the center of mass (referred to as ``center migration'').
\end{enumerate}
The volume filling factor of the generated aggregate depends on the selection mechanism. For a given coordination number, aggregates generated with ``shortest migration'' feature the lowest filling factor whereas the ``center migration'' method leads to the most compact aggregates.

The relation between the filling factor and the coordination number is displayed in Fig.\,\ref{fig:ff_nc}, for the different preparation methods. For comparison we show the results for the hexagonal close packing as well, see \citet{2013A&A...551A..65S} for more details. After generating a sufficiently large aggregate a cuboidal shaped dust cake of the desired size is sliced out to be used for the subsequent numerical experiments.

\section{Tensile strength}

\subsection{Setup}\label{sec:tensile_strength_setup}

\begin{figure}
\resizebox{\hsize}{!}{\includegraphics{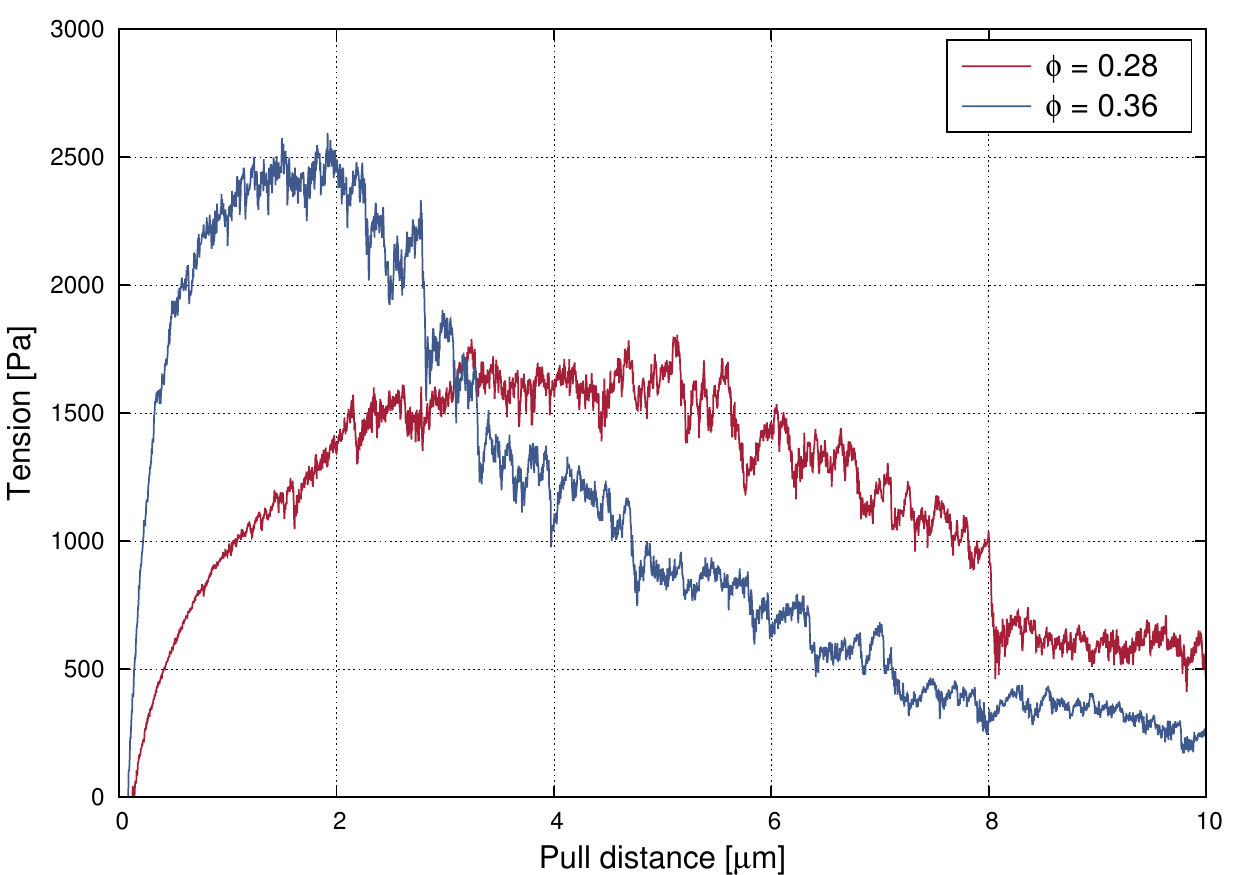}}
\caption{Example of the tension required to pull the plates apart from each other for two cubical samples of different porosity. As the pull distance increases the required force decreases because cracks form in the sample.}
\label{fig:tensile_strength_example} 
\end{figure}

\begin{figure*}
\resizebox{\hsize}{!}{\includegraphics{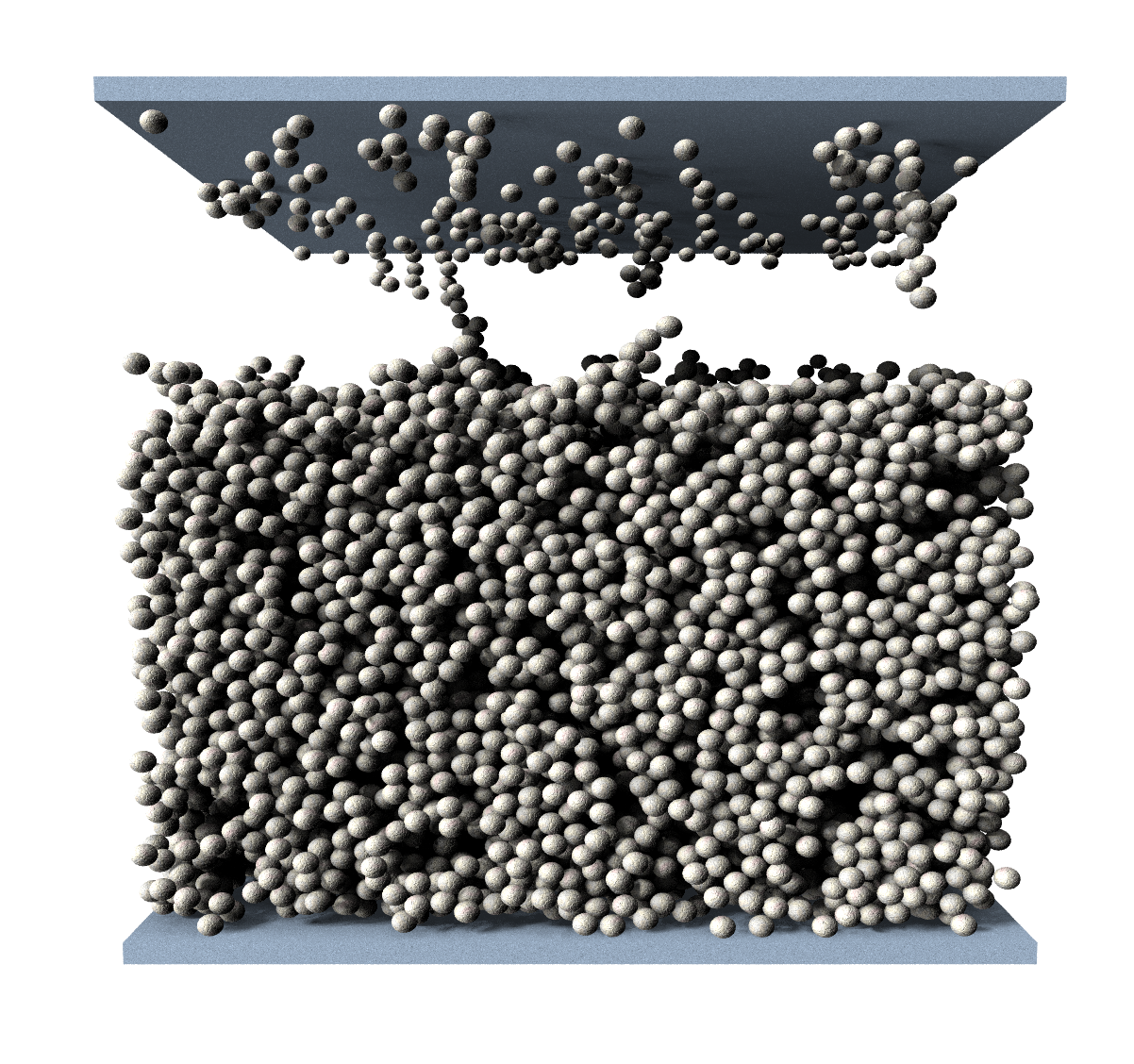} \hfill \includegraphics{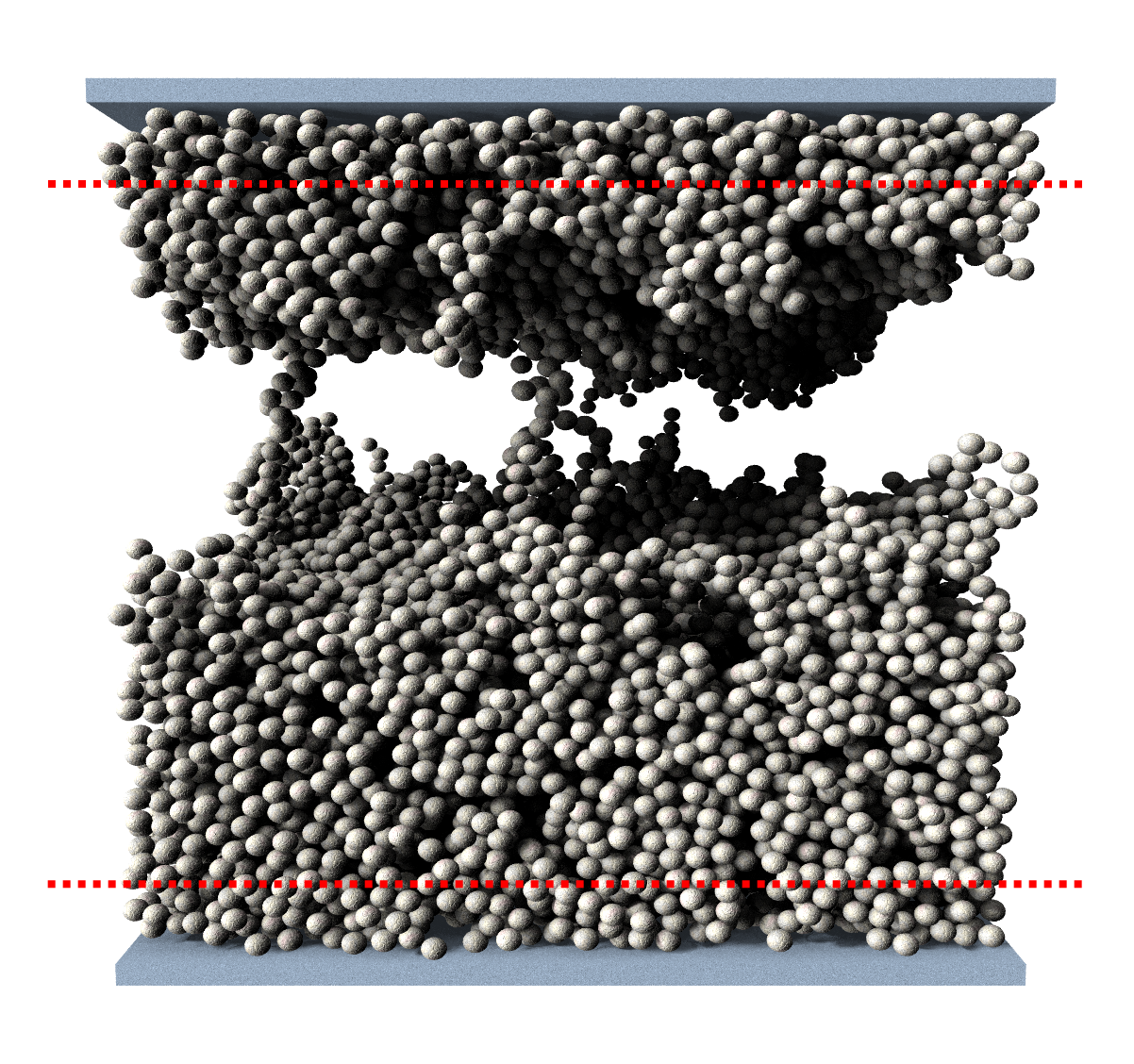}}
\caption{Outcome of a typical pull experiment on a cubical sample agglomerate. \textit{Left:} Since the adhesion between particle-wall contacts is stronger than between particles the uppermost layer of particles is ripped off when pulling the plates away from each other. \textit{Right:} Adhesion between particles that are close to one of the plates has been artificially increased. The red dotted line indicates where the additional gluing effect sets in.}
\label{fig:wall_glue}
\end{figure*}

In principle, the procedure to determine the tensile strength of a given sample is simple: After attaching two plates at the top and bottom of the cubic aggregate the plates are pulled apart at constant speed. During this process the pulling force exerted on these plates is measured. At first, the force will increase with the distance between the plates. If a certain distance is exceeded, cracks will form. Thus, the strength of the sample is reduced and the force required to pull the two plates apart decreases. An example showing a typical relation between the pull distance and tension is shown in Fig.\,\ref{fig:tensile_strength_example}. It is very similar to experimental data \citep[see][Fig.\,4]{PhysRevLett.93.115503}. 

The force is determined by summing up the individual interaction forces of all monomers that are in contact with the wall. In case of the tensile strength, only the component in the direction of the motion of the wall (which is equivalent to the normal vector of the wall) is taken into account. To allow easier comparison between different sized samples we use their base area to normalize the values and plot the corresponding tension instead of the force. 

In accordance with \citet{2006ApJ...652.1768B}, we define the maximum tension that is measured during a run as the tensile strength of the sample. The displacement at which force/pressure peaks depends on the porosity of the sample. In a sample with high filling factor monomers are fixed tightly, which hampers internal restructuring. In contrast, in a fluffy sample individual chains of monomers can be unfolded and thus the material can be stretched out significantly before the formation of cracks sets in.

To model this setup in our simulations, the sample is put into a box of flat walls. Before slowly moving the top and bottom wall away from each other we must ensure that a sufficient number of monomers is in contact with these walls. For this purpose the top and bottom wall are slowly pushed inwards a short distance. For samples with filling factors below $\phi = 0.2$ we use a value of one monomer radius whereas for more compact samples we decrease the distance to $0.5\,-\,0.1$ monomer radii.

When pulling the two plates away from each other another problem arises: The critical force $F_\mathrm{c}$ required to break a contact between two monomers is given by $F_\mathrm{c} = 3 \pi \gamma R$, where $\gamma$ denotes the surface energy and $R$ the reduced radius of the two particles \citep{1971RSPSA.324..301J}. The wall is modeled as a particle of infinite radius, which means that the reduced radius of a particle-wall contact equals twice the reduced radius of a particle-particle contact \citep{2012A&A...541A..59S}. Thus, contacts between two particles can be broken more easily than particle-wall contacts. As a result, the monomers that are in contact with one of the plates tend to get ripped off the remaining sample (see left panel of Fig.\,\ref{fig:wall_glue}).

To counter this effect we artificially increase the strength of the adhesion between two monomers dependent on the distance to the plates. To achieve this ``gluing effect'' the force/potential of the normal interaction that is responsible for the adhesion is multiplied with a gluing factor $\kappa$. To avoid discontinuities in the particle interaction $\kappa$ is interpolated linearly depending on the distance to the closest plate. Above a threshold of 8 particle radii $\kappa$ is set to 1 and thus the default JKR interaction is used. As shown in the right panel of Fig\,\ref{fig:wall_glue}, this mechanism leads to the rupture somewhere in the middle of the sample rather than just tearing off the upper- or lowermost layer of monomers. As the first cracks will form where the aggregate is weakest, the exact location is random owing to the inhomogeneous structure of the aggregate.

\begin{figure}
\resizebox{\hsize}{!}{\includegraphics{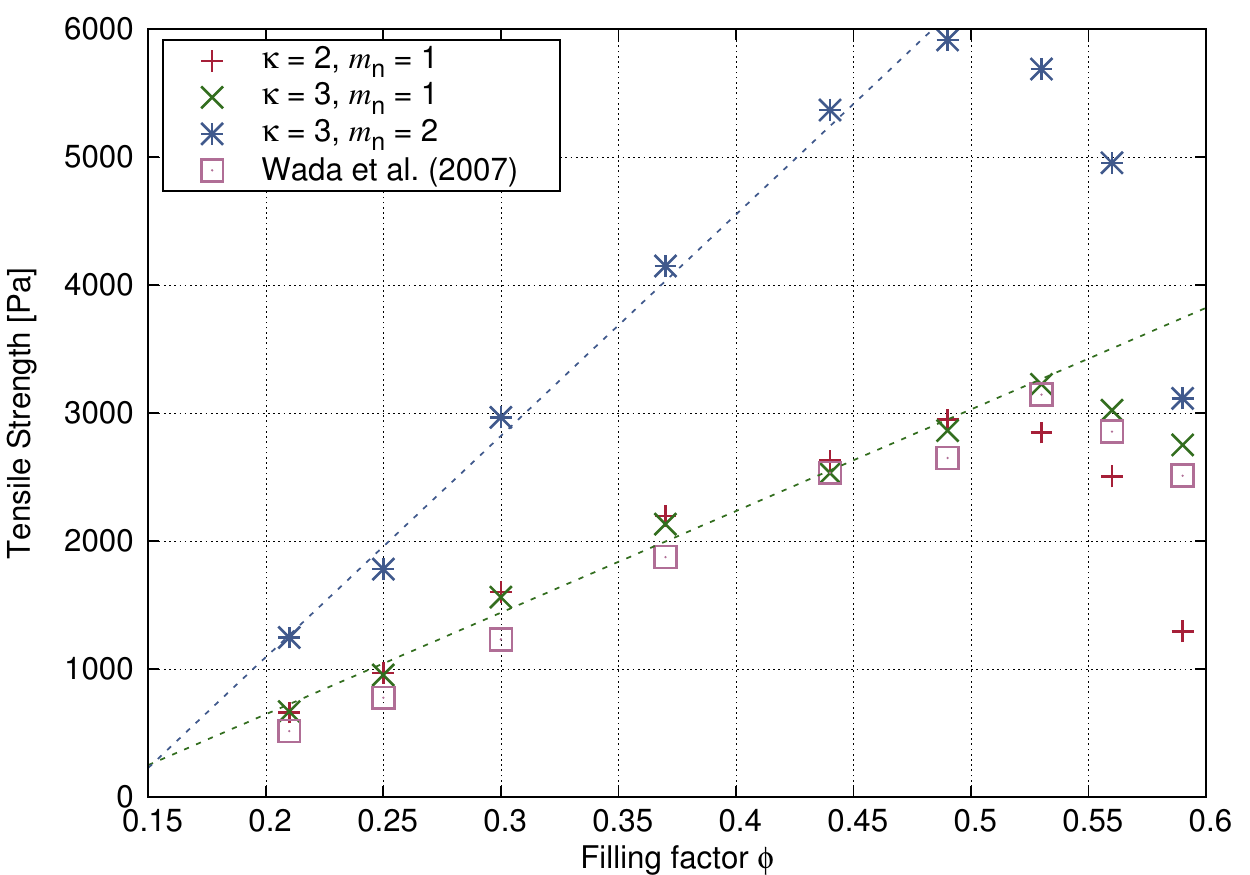}}
\caption{Tensile strength for different wall glue factors $\kappa$ and normal interaction modifiers $m_\mathrm{n}$ using the model from \citet{2012A&A...541A..59S} ($m_\mathrm{r} = 8$, $m_\mathrm{s} = 2.5$). The dotted lines represent linear fits for filling factors below $0.5$. For comparison, we also performed simulations using the model employed by \citet{2007ApJ...661..320W} ($m_\mathrm{r} = m_\mathrm{s} = 1$). All samples are BAM cakes with an edge length of $50 \times 50 \times 30\,\mathrm{\mu m}$.}
\label{fig:tensile_strength_interaction}
\end{figure}

We tested different maximum values of $\kappa$ and found that a value of 2 is sufficient for our purpose. For $\kappa < 1.5$, samples do not break in the center anymore. On the other hand, larger values do not alter the measured tensile strength significantly (see Fig.\,\ref{fig:tensile_strength_interaction}).

\subsection{Results}\label{sec:tensile_strength_results}
Apart from the wall gluing factor $\kappa$ there are several other parameters whose influence has to be studied. To determine the influence of the rolling and sliding interaction we performed a series of simulations using $m_\mathrm{r} = m_\mathrm{s} = 1$, which is equivalent to the model of \citet{2007ApJ...661..320W}. In this and all of the following simulations we used a wall gluing factor of $\kappa = 2$. Apparently, internal restructuring which is governed by rolling and sliding does not play a major role when determining the tensile strength (see the purple squares in Fig.\,\ref{fig:tensile_strength_interaction} showing results for the model from \citet{2007ApJ...661..320W}).

While the sample aggregate is torn apart contacts between the monomers have to be broken. Therefore, we expect that the measured tensile strength depends on the number of contacts that have to be broken as well as the critical force $F_\mathrm{c}$ which is necessary to break a contact between individual monomers. To check this hypothesis we alter the strength of the normal force by multiplying it with the normal force modifier $m_\mathrm{n}$. Indeed, when doubling the strength of the normal interaction (and thus $F_\mathrm{c}$) by setting $m_\mathrm{n} = 2$, we observe a steeper increase of the tensile strength with the filling factor (blue asterisks in Fig.\,\ref{fig:tensile_strength_interaction}). When determining a linear fit for filling factors below $0.5$ to the $\kappa = 3, m_\mathrm{n} = 1$ and $\kappa = 3, m_\mathrm{n} = 2$ simulations we get a slope of $7.9\,\mathrm{kPa}$ and $15.3\,\mathrm{kPa}$, respectively. Their ratio of $15.3 / 7.9 = 1.94$ is very close to the value of the normal interaction modifier $m_\mathrm{n} = 2$. This strongly suggests that the pull off force $F_\mathrm{c}$ is critical for the measured value of the tensile strength.

\begin{figure}
\resizebox{\hsize}{!}{\includegraphics{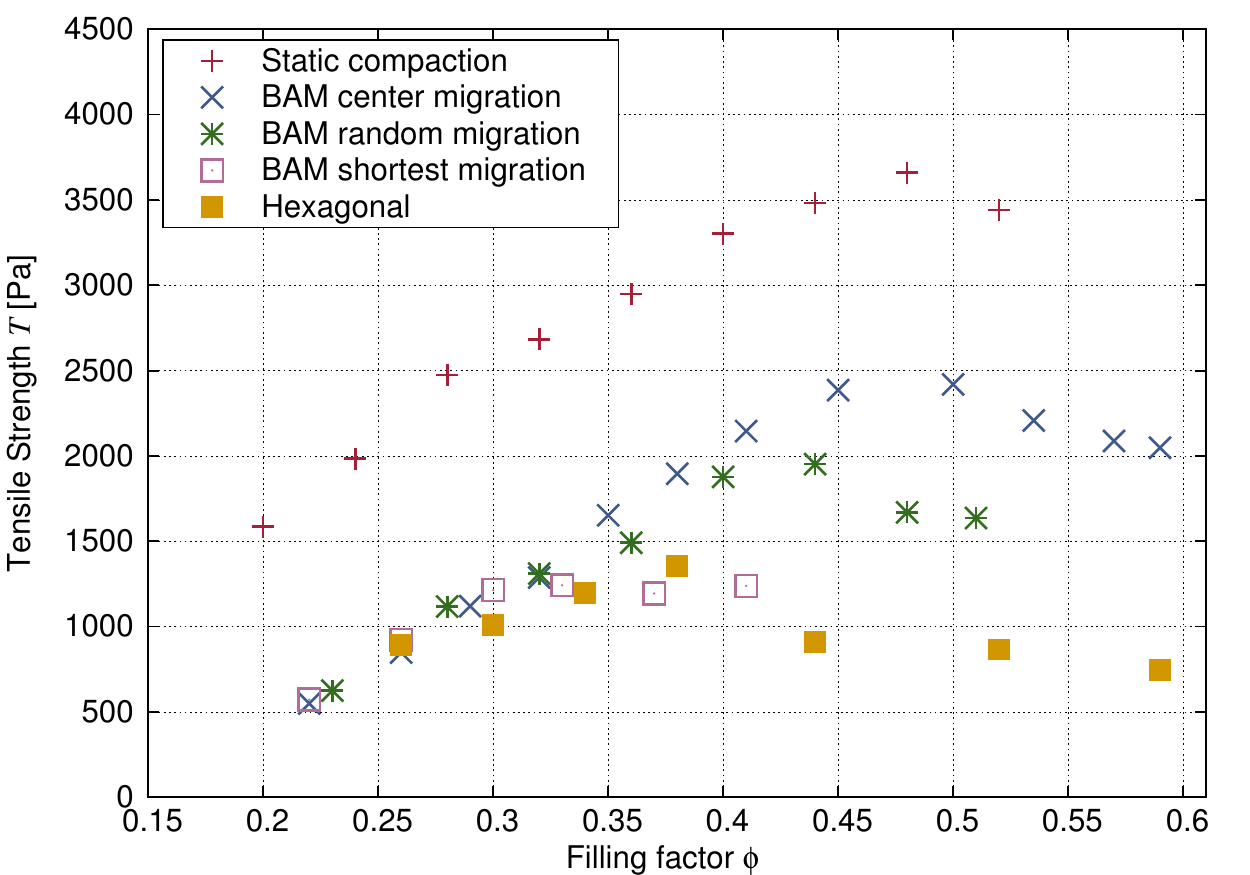}}
\caption{Comparison of the relation between the filling factor $\phi$ and the tensile strength of different sample types. Most noticeably, the measured tensile strength always drops when a certain type-specific filling factor is exceeded.}
\label{fig:tensile_strength_cake_types}
\end{figure}

Independent of $\kappa$, $m_\mathrm{n}$, or the rolling and sliding interaction there is a striking drop of the tensile strength for filling factors above $0.5$. To unravel its cause we first used a different type of samples. Much to our surprise, the critical filling factor at which the tensile strength drops is different for each type of aggregate and close to the maximum filling factor that may be achieved by a given generation method (see Fig.\,\ref{fig:tensile_strength_cake_types}). Apparently, the micro-mechanical behavior is not governed by the volume filling factor alone.

\begin{figure}
\resizebox{\hsize}{!}{\includegraphics{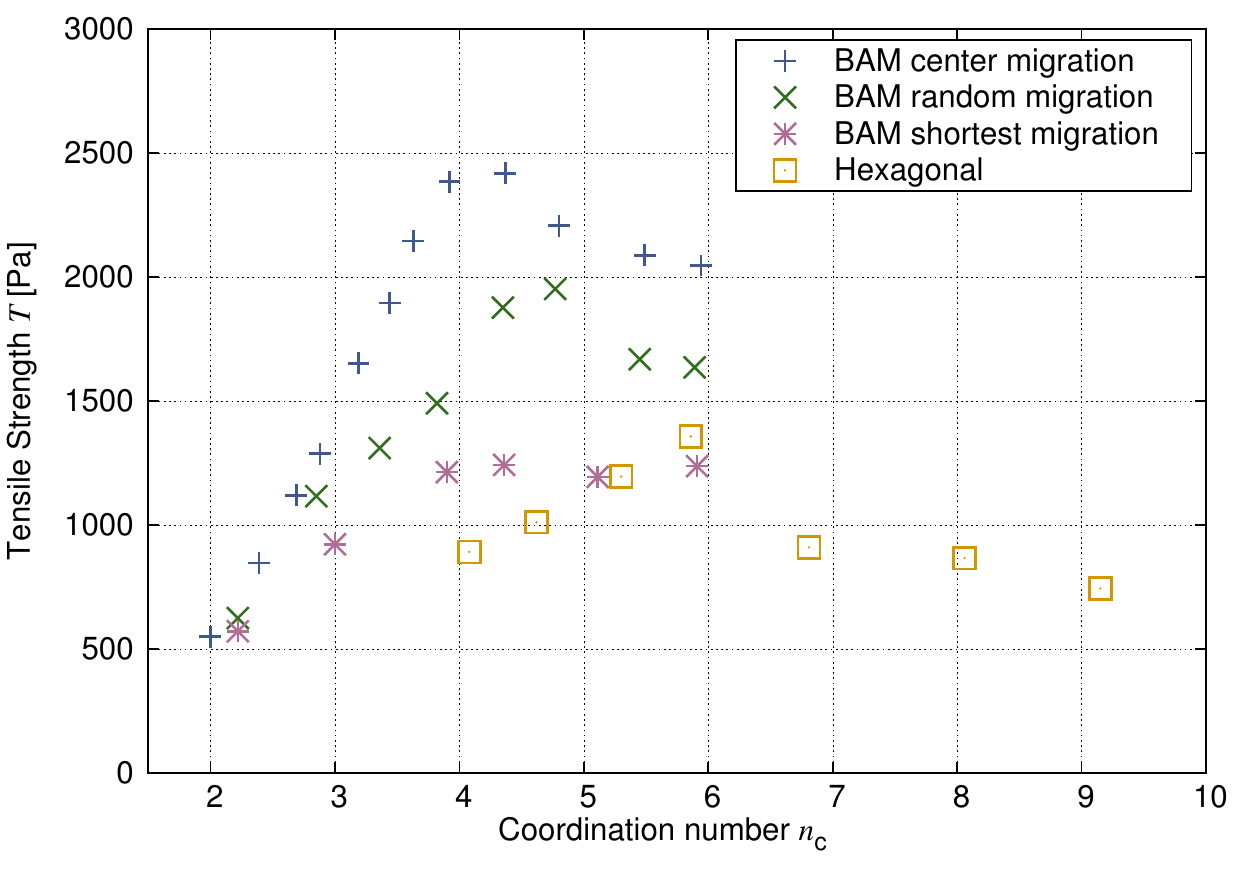}}
\caption{Comparison of the relation between tensile strength and initial coordination number $n_\mathrm{c}$ of different sample types. The tensile strength of the different BAM aggregates drops when for values of $n_\mathrm{c}$ around 4.5. For the hexagonal close packing aggregates we see a clear drop for $n_\mathrm{c} \rightarrow 6$.}
\label{fig:tensile_strength_cake_types_nc}
\end{figure}

Therefore we take the average coordination number $n_\mathrm{c}$ into account. From the relation between the filling factor and the coordination number of the different sample types (see Fig.\,\ref{fig:ff_nc}) we find that the drop of the measured tensile strength of BAM aggregates coincides with a value of $n_\mathrm{c}$ around $\approx 4.5$ (see Fig.\,\ref{fig:tensile_strength_cake_types_nc}). This points to the influence of the coordination number on the micro-mechanical properties of the sample aggregates. If the average coordination number is low, the majority of monomers may react to an external stress by rearranging themselves. Thus, a large number of monomers participate in absorbing the external stress and the aggregate exhibits rather ductile behavior. As the coordination number increases, monomers are fixated in their positions more tightly. For compact aggregates the monomers cannot rearrange themselves freely anymore which means that a lower number of monomers has to absorb the applied strain. Therefore, the aggregates become brittle.

The effect of brittleness can clearly be seen in case of the hexagonal lattice aggregates. Because of their regular, crystal like structure \citep[see][Fig.\,1a]{2013A&A...551A..65S}, their capability of internal restructuring is very limited. Thus, contacts break very easily when external strain is applied. As a result, the measured tensile strength is considerably lower compared to BAM or static compaction aggregates.

Additionally, the pressure exerted on the sample when slowly pressing on the top and bottom walls suffices to disrupt very compact aggregates. Moving the top and bottom walls inwards by a distance of $0.5$ to $0.15$ particle radii is necessary to establish a firm contact between the walls and the sample. E.g.\,in the case of a BAM (center migration) aggregate with $\phi = 0.59$ the average coordination number decreased from $5.94$ to $4.95$ after moving the top wall inwards by a distance of only $0.1$ monomer radii. This means that the strength of the sample is lowered during the preparation process. This raises the question whether the transition from ductile to brittle behavior or the disturbance when affixing the plates is the dominant effect.

Recently, \citet{2013A&A...554A...4K} presented a different approach to determine the compressive strength of highly porous ($\phi < 0.1$) dust aggregates by using periodic boundary conditions. A similar approach might work for the tensile strength as well and would avoid the problem of attaching the plates onto a highly compact sample without lowering its strength. Luckily, measuring the tensile strength of hexagonal lattice aggregates also allows us to circumvent this problem. Because of their regular grid structure the contact between the wall and all particles of the top/bottom layer is established without compacting the sample. Nevertheless, we observe a significant drop of the tensile strength for $n_\mathrm{c} \rightarrow 6$ (see Fig.\,\ref{fig:tensile_strength_cake_types_nc}). This observation allows us to conclude that the disruption caused by affixing the plates only plays a secondary role.

\begin{figure}
\resizebox{\hsize}{!}{\includegraphics{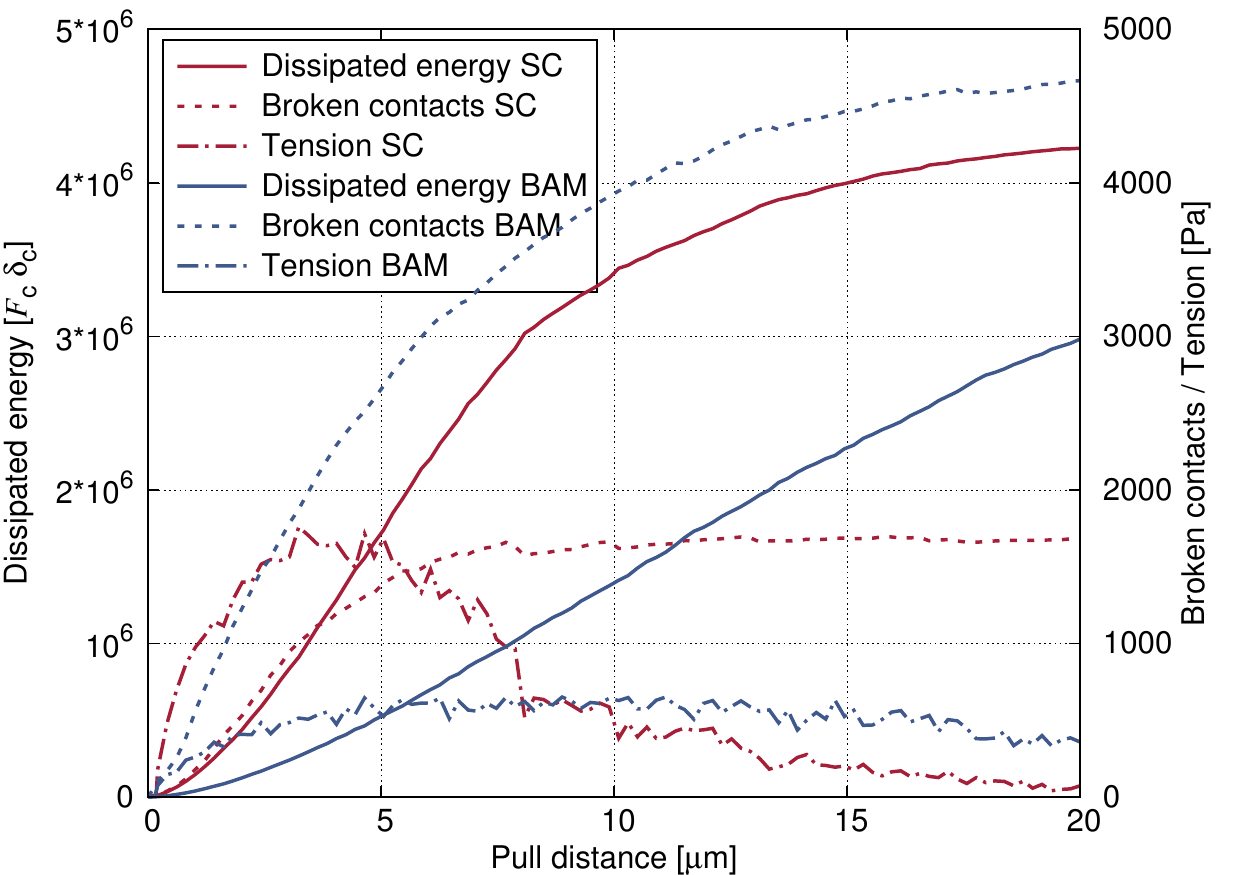}}
\caption{The dissipated energy (solid line) and the total number of broken contacts (dotted line) of BAM and static compaction (SC) aggregates while pulling the plates apart. The size of both samples is $50\times50\times30\,\mathrm{\mu m}$ and their initial filling factor is $\phi = 0.28$. The critical force $F_\mathrm{c}$ and critical distance $\delta_\mathrm{c}$ at which a contact breaks are used to normalize the energies. For comparison, the corresponding tension (dashed-dotted line) is also plotted.}
\label{fig:dissipated_energy}
\end{figure}

To explain the discrepancy between BAM and static compaction aggregates as shown in Fig.\,\ref{fig:tensile_strength_cake_types}, the process of pulling the plates apart is analyzed more closely. Comparing the dissipated energy of the different types of aggregates we clearly see that tearing apart the static compaction aggregate requires more energy (solid lines in Fig.\,\ref{fig:dissipated_energy}). Thus, a higher value of the tensile strength is measured. Tracking the number of contacts that have been broken since the start of the simulation provides us with a hint to an explanation of this observation: Compared to the BAM aggregate, less than half as many contacts break while pulling apart the static compaction aggregate (dotted line in Fig.\,\ref{fig:dissipated_energy}) although more kinetic energy is dissipated. 

This observation might be surprising at first sight, yet it illustrates the importance of the internal structure for the outcome of the measurements. In reaction to the applied strain the internal structure of the aggregates changes, where inelastic rolling accounts for $\approx 80\%$ and the breaking of contacts for only $\approx 3\%$ of the total dissipated energy for both aggregate types. In case of the static compaction aggregate the final number of broken contacts is reached much more quickly after pulling the sample roughly $7.5\,\mathrm{\mu m}$ apart. Likewise, the peak of the measured tension is reached earlier. At this point, a lot more energy has been dissipated by internal restructuring (mainly by inelastic rolling) as compared to the BAM aggregate.

However, for the BAM aggregates the total number of broken contacts increases much faster. This means that the internal structure of this type of aggregate allows for less restructuring before contacts start to break. As roughly $97\%$ of the dissipated energy are required for restructuring rather than breaking of the contacts we measure a lower tensile strength for BAM aggregates.

We think that the reason behind this observation lies in the fact that the tensile strength test is the reversal of the compression process by which the static compaction aggregates have been generated. While being slowly compacted to the desired filling factor sufficient time had been given to the fractal chains of the initial RBD aggregates to rearrange themselves. As a result, the monomers of the static compaction aggregates adopt a structure that is favorable to withstand an external load.

While pulling the plates apart the total number of broken contacts (dotted line in Fig.\,\ref{fig:dissipated_energy}) sometimes decreases. This happens when the connection between two already stretched out parts of the sample breaks and the both parts ``snap back''.

We also checked whether there are any preferential directions resulting from the generation process. For this reason we rotated cubic static compaction samples by $90^{\circ}$ before determining the tensile strength. Reassuringly, we measured the same values (not shown in this work) and may rule out that the direction of the compaction induces any preferred direction in the structure of the sample.

\begin{figure}
\resizebox{\hsize}{!}{\includegraphics{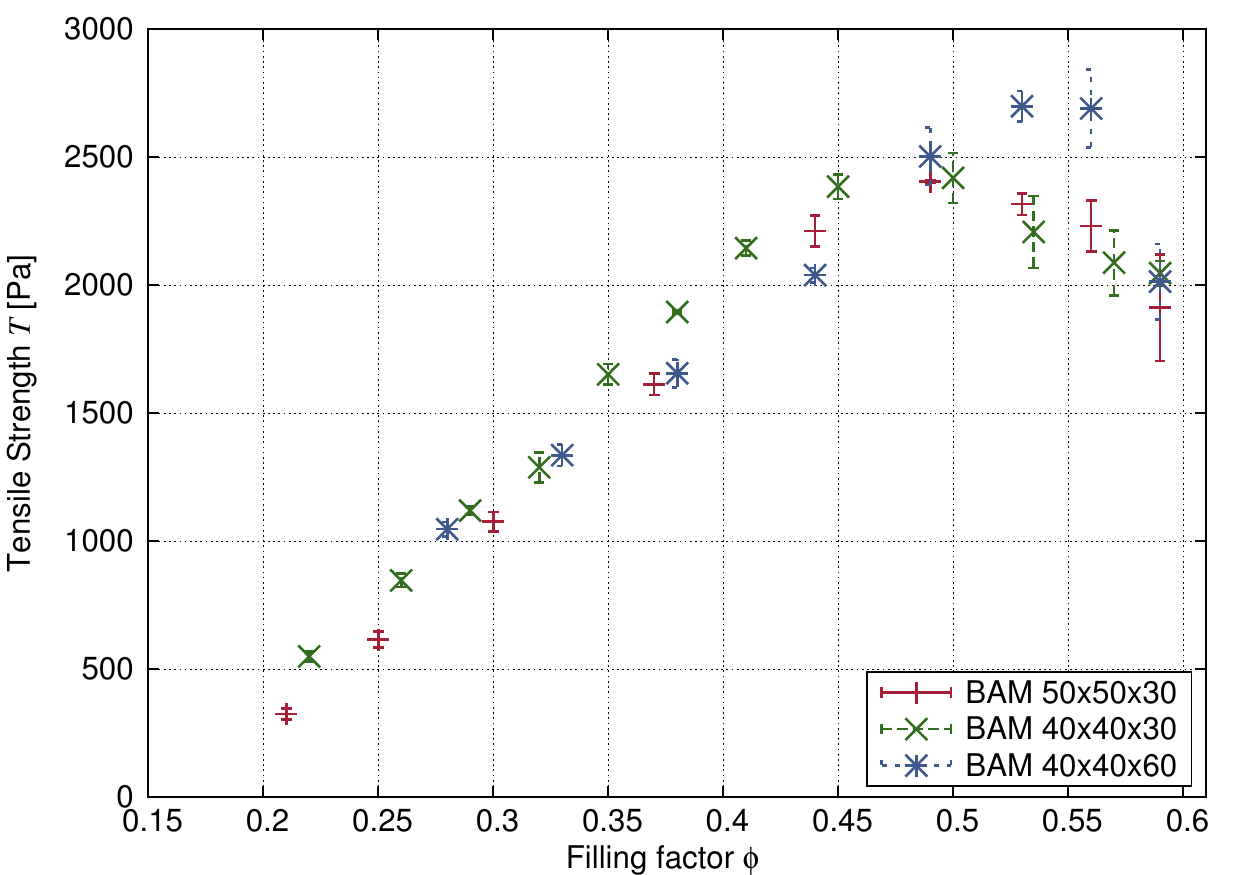}}
\caption{Tensile strength measurements for BAM (center migration) aggregates of different size. The error bars have been determined by averaging the results of six runs with different samples.}
\label{fig:tensile_strength_cake_dim}
\end{figure}

\begin{figure}
\resizebox{\hsize}{!}{\includegraphics{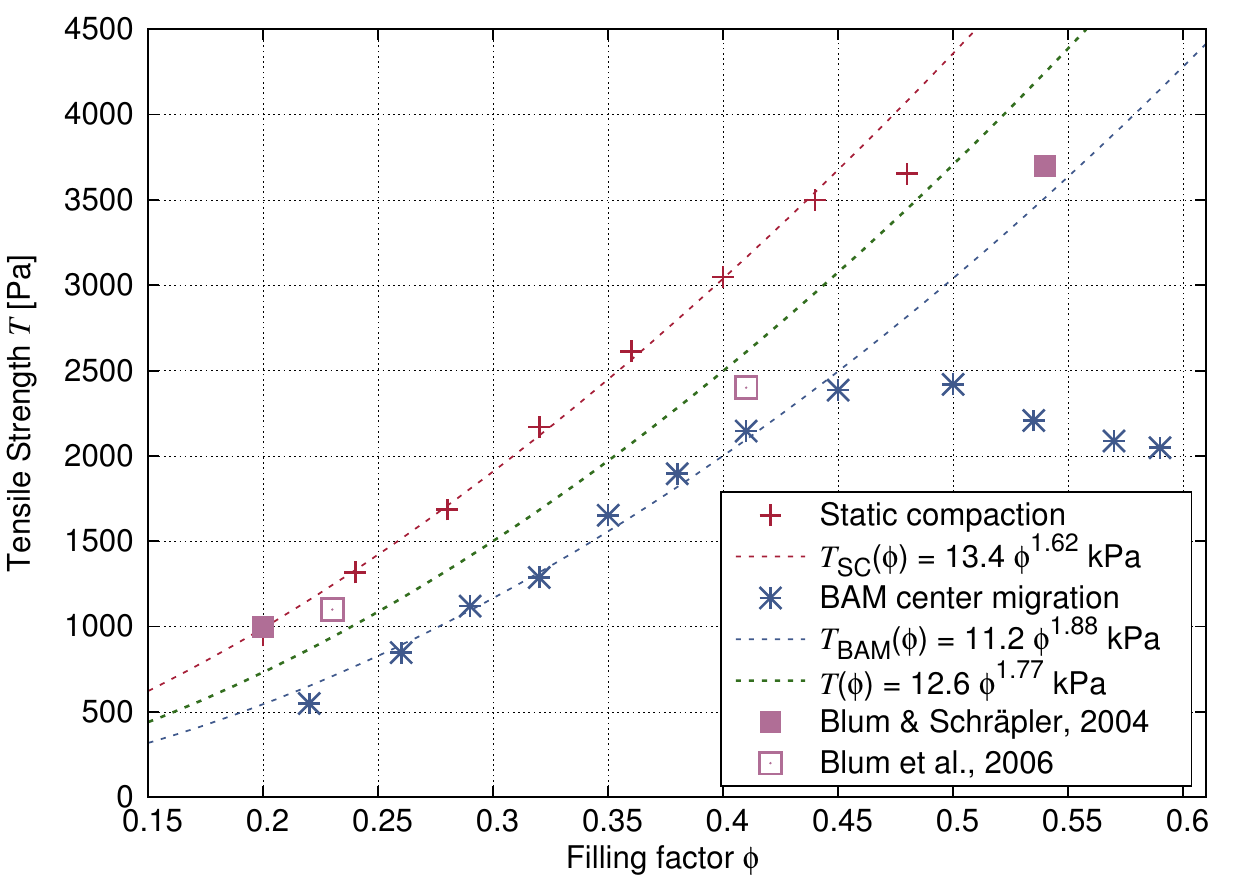}}
\caption{Comparison of the results of simulations with static compaction and BAM (center migration) aggregates to laboratory experiments by \citet{PhysRevLett.93.115503, 2006ApJ...652.1768B}. The fitting curve has been obtained by combining both data sets for filling factors $\phi < 0.5$.}
\label{fig:tensile_strength_fit}
\end{figure}

As a last step we varied the geometry of the samples to check if the size influences the results. As it can be seen in Fig.\,\ref{fig:tensile_strength_cake_dim} the results do not vary significantly if we alter the size of the samples. In Fig.\,\ref{fig:tensile_strength_fit} we compare our results with laboratory experiments performed by \citet{PhysRevLett.93.115503} and \citet{2006ApJ...652.1768B}. Taking into account that their samples have been produced by static compaction our results show good agreement with their data for filling factors below $\phi = 0.5$. Because of the drop of the tensile strength for $n_\mathrm{c} \rightarrow 6$ explained in the previous paragraphs we cannot compare our simulations to laboratory experiments for higher filling factors. Earlier compression simulations already indicated that our physical model may not be valid for highly compact aggregates anymore \citep{2012A&A...541A..59S}. Luckily, the filling factor regime relevant for the growth processes of planetesimals is below $0.5$ \citep[e.g.\ ][]{2011ApJ...742....5T}. 

To determine the fitting curve depicted in Fig.\,\ref{fig:tensile_strength_fit} only data points for filling factors below 0.5 have been taken into account. Because of the significant difference between the static compaction and BAM aggregates we determined two fit curves $T_\mathrm{SC}(\phi)$ and $T_\mathrm{BAM}(\phi)$, respectively. We obtain
\begin{eqnarray}T_\mathrm{SC}(\phi) = 13.4\,\phi^{1.62}\,\mathrm{kPa},\label{eq:ts_fit_sc}\end{eqnarray}
and
\begin{eqnarray}T_\mathrm{BAM}(\phi) = 11.2\,\phi^{1.88}\,\mathrm{kPa}.\label{eq:ts_fit_bam}\end{eqnarray}

Based on their generation process the laboratory samples should resemble the static compaction cakes. Indeed, for $\phi \approx 0.2$ our simulations agree very well with laboratory experiments \citep{PhysRevLett.93.115503, 2006ApJ...652.1768B}. However, for higher filling factors the laboratory results lie somewhere between the static compaction and BAM results. In private conversation J\"urgen Blum (Braunschweig) pointed out that creating more compact samples in the lab sometimes proved to be a difficult task. Thus, we also determined a fit $T(\phi)$ to the combined results of the static compaction and BAM aggregates. Using the combined data points from both aggregate types shown in Fig.\,\ref{fig:tensile_strength_fit} for a single fit $T(\phi)$, we find for values of $\phi < 0.5$
\begin{eqnarray}T(\phi) = 12.6\,\phi^{1.77}\,\mathrm{kPa}.\label{eq:ts_fit}\end{eqnarray}

\subsection{Influence of the monomer size}

\begin{figure}
\resizebox{\hsize}{!}{\includegraphics{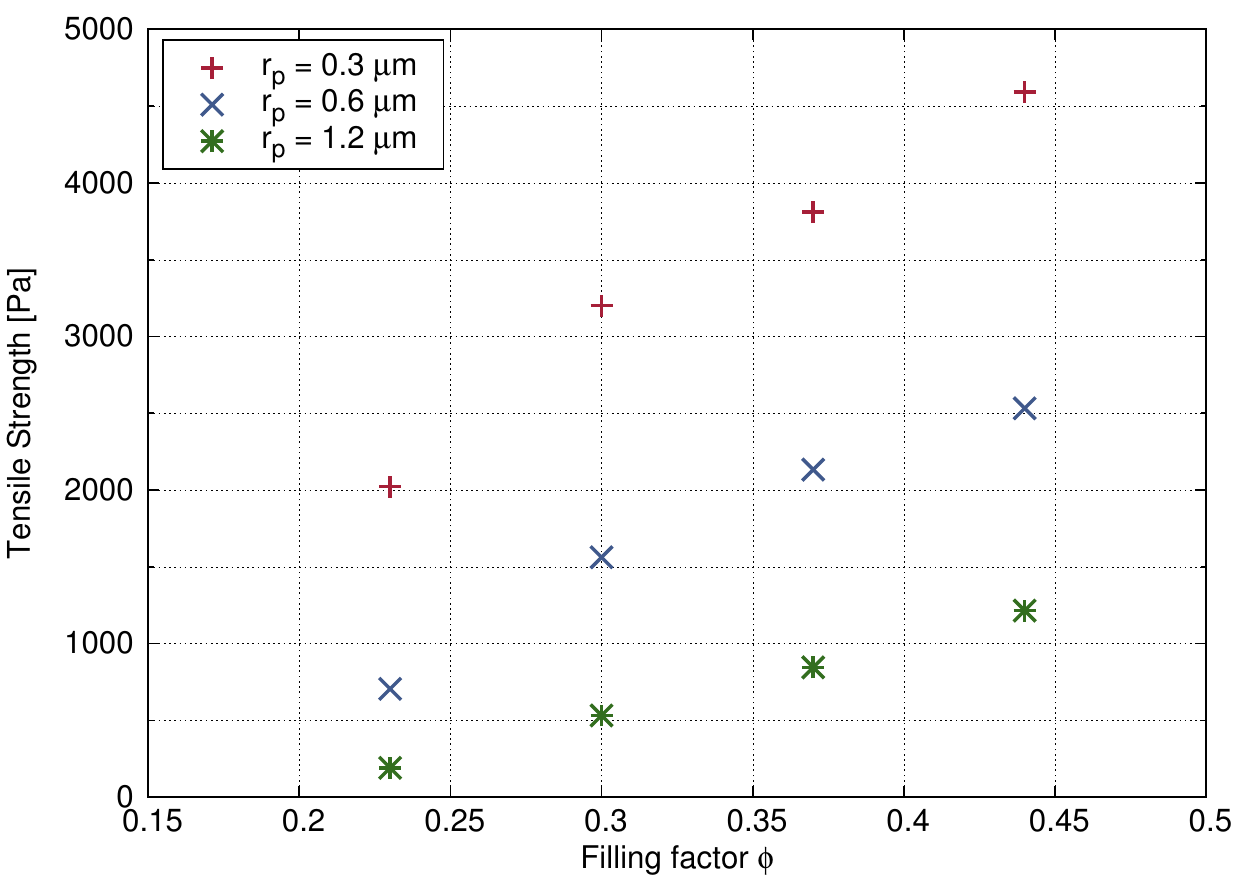}}
\caption{Tensile strength for aggregates composed of differently sized monomers. The default monomer radius is $r_\mathrm{p} = 0.6\,\mathrm{\mu m}$, which is also used in the corresponding laboratory experiments.}
\label{fig:tensile_strength_monomer_size}
\end{figure}

Most other numerical simulations dealing with Silicates have been performed with monomers with a diameter of $1.2$ and $1.5\,\mathrm{\mu m}$ because these sizes have been used in laboratory experiments with spherical Silicate grains. Out of curiosity we varied the size of the monomers. In general, it can be said that according to JKR-theory the adhesion forces increase as grains get smaller. Indeed, we find that the tensile strength depends strongly on the size of the monomers (see Fig.\,\ref{fig:tensile_strength_monomer_size}). Note that our interaction model has not been calibrated for monomer radii other than $r_\mathrm{p} = 0.6\,\mathrm{\mu m}$. Therefore the rolling and sliding modifiers $m_\mathrm{r}$ and $m_\mathrm{s}$ may not have the correct values to properly describe restructuring processes. However, in case of the tensile strength this problem does not arise as we have already seen that it is mainly governed by the normal interaction (see Fig.\,\ref{fig:tensile_strength_interaction}).

From Fig.\,\ref{fig:tensile_strength_interaction} we see that the tensile strength scales linearly with the normal interaction. Altering the monomer size by a factor of 2, for the transition from $0.6\,\mathrm{\mu m}$ to $1.2\,\mathrm{\mu m}$ sized monomers the change of the measured tensile strength differs from 2. At first glance this may seem odd as the critical pull off force $F_\mathrm{c}$ depends linearly on the monomer radius. However, the dependence of the normal force acting upon the monomers before they are separated on the monomer radius is non-linear \citep[see][Eqs.\,2 and 3]{2012A&A...541A..59S}. Nevertheless, these simulations confirm the importance of the pull off force $F_\mathrm{c}$ that has already been shown in Fig.\,\ref{fig:tensile_strength_interaction}.

The results clearly demonstrate the effect of the stickiness of the single monomers on the tensile strength. For future work it would be interesting to perform simulations with aggregates composed of differently sized monomers.

\section{Shear strength}

\subsection{Setup}\label{sec:shear_strength_setup}
\begin{figure}
\resizebox{\hsize}{!}{\includegraphics{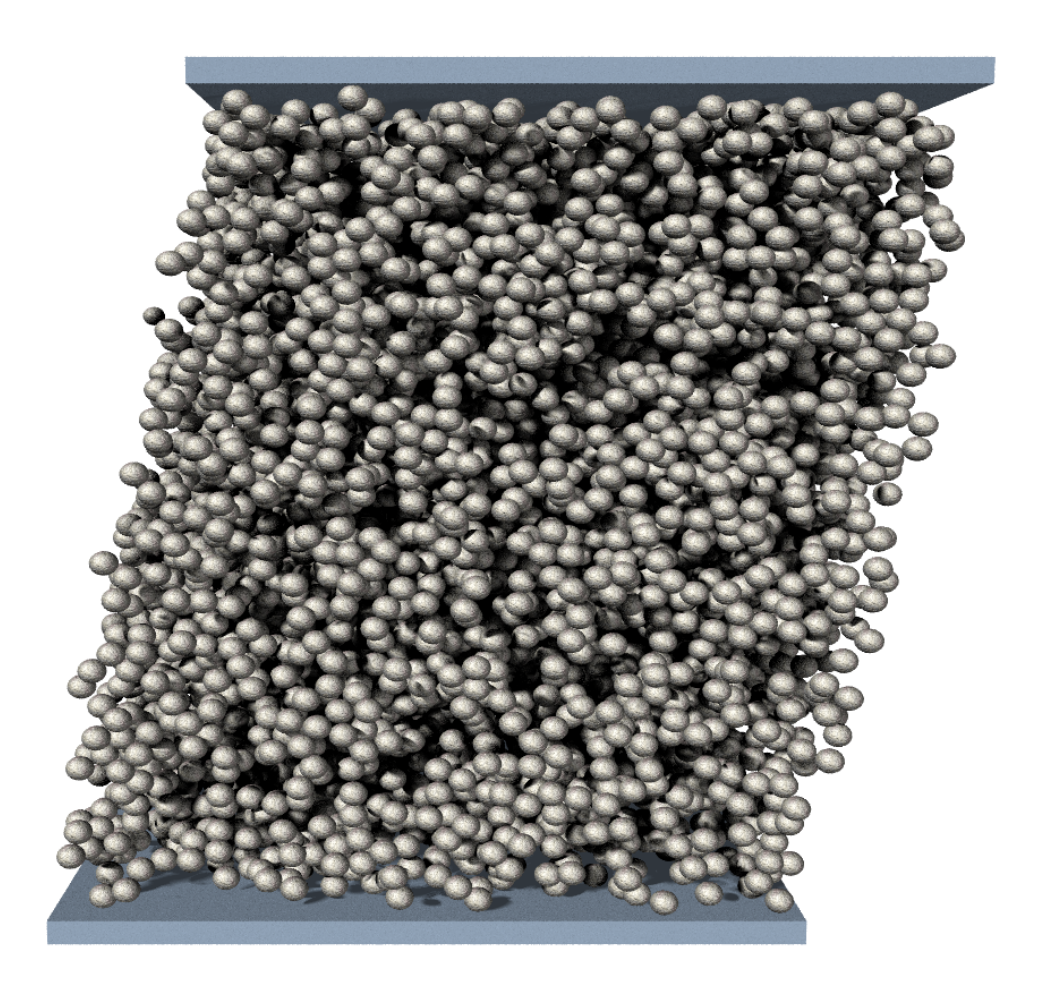}}
\caption{A snapshot taken during a shear strength test using an aggregate with an edge length of $30 \times 30 \times 40\,\mathrm{\mu m}$}. The upper plate is slowly moving to the right. Adhesion between particles that are close to one of the plates has been artificially increased.
\label{fig:shear_strength_setup}
\end{figure}

\begin{figure}
\resizebox{\hsize}{!}{\includegraphics{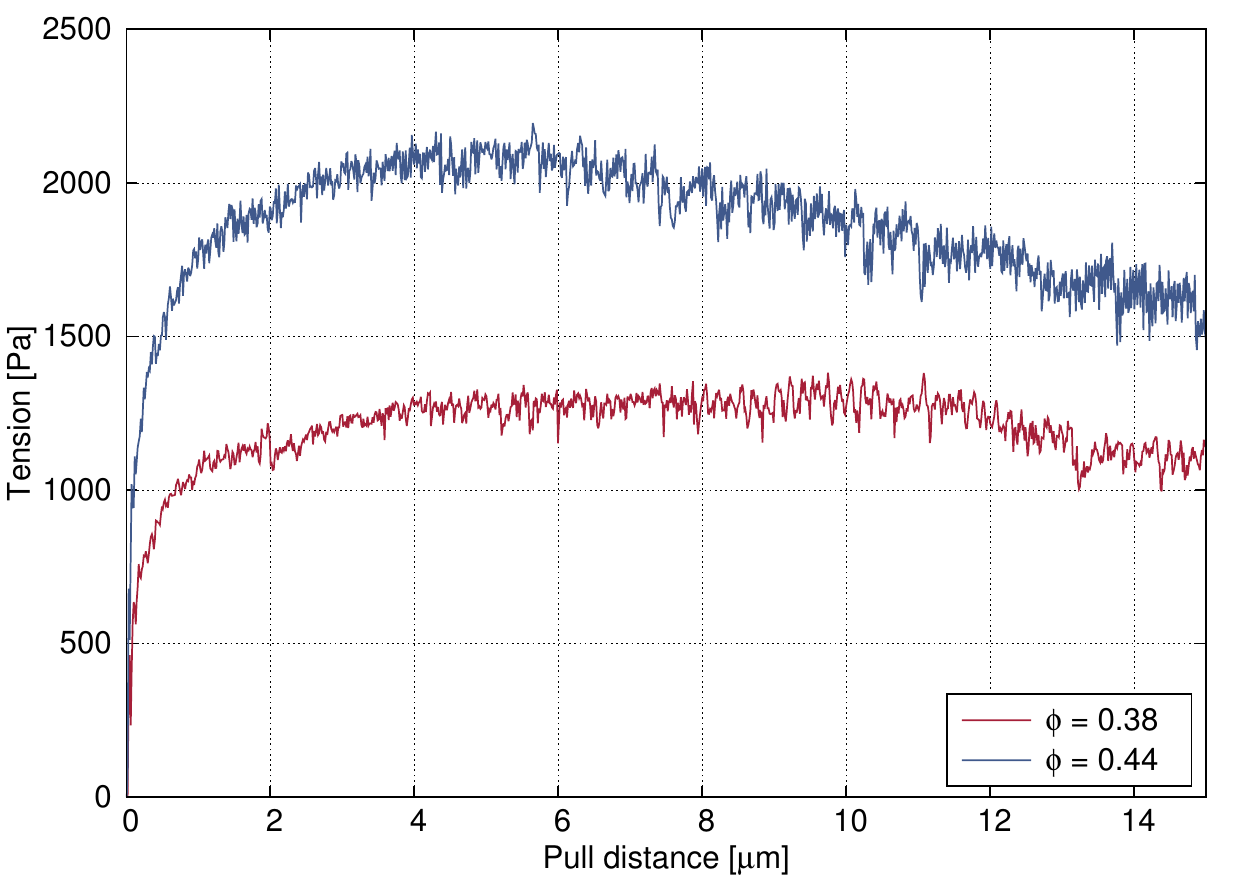}}
\caption{Example of the tension (force per area) acting on the upper plate during the shearing motion.}
\label{fig:shear_strength_example}
\end{figure}

The shear strength of a porous aggregate is determined in a similar way as the tensile strength. As before, two plates are attached to the top and bottom of the sample. During the shearing motion of the plates the force acting on them is tracked. However, in case of the shear strength the direction of motion is perpendicular to the normal of the wall, i.e. tangential to a cuboid surface. During the simulation the vertical positions of the top and bottom wall remain fixed to keep the filling factor constant. This means, similar to the ``fixed walls'' used in the work of \citet{2012A&A...541A..59S} the normal component of the force acting upon the walls is ignored. As before, the initial base area of the sample is used to normalize the force. 

The setup closely resembles the tensile strength setup. A firm contact between the sample and the plates is achieved in the same way as described in Sec.\,\ref{sec:tensile_strength_setup}. To prevent the monomers that are in contact with the moving wall from being torn away from the sample, an additional ``gluing effect'' is applied to particles that are close to one of the plates. A snapshot taken during a typical simulation is depicted in Fig.\,\ref{fig:shear_strength_setup}.

As the top plate is slowly moving shearing sets in. With increasing pulling distance cracks will form and reduce the strength of the sample. Thus, we expect a similar shape as for the tensile strength if we plot the tension acting on the moving plate with respect to the displacement. Indeed, the example shear strength curve shown in Fig.\,\ref{fig:shear_strength_example} resembles the curves shown in Fig.\,\ref{fig:tensile_strength_example}.

Similar to the tensile strength case, we define the shear strength as the maximum tension that is measured during the simulation. Again, the higher the porosity of a sample the larger the necessary displacement at which the force peaks.

\subsection{Results}

\begin{figure}
\resizebox{\hsize}{!}{\includegraphics{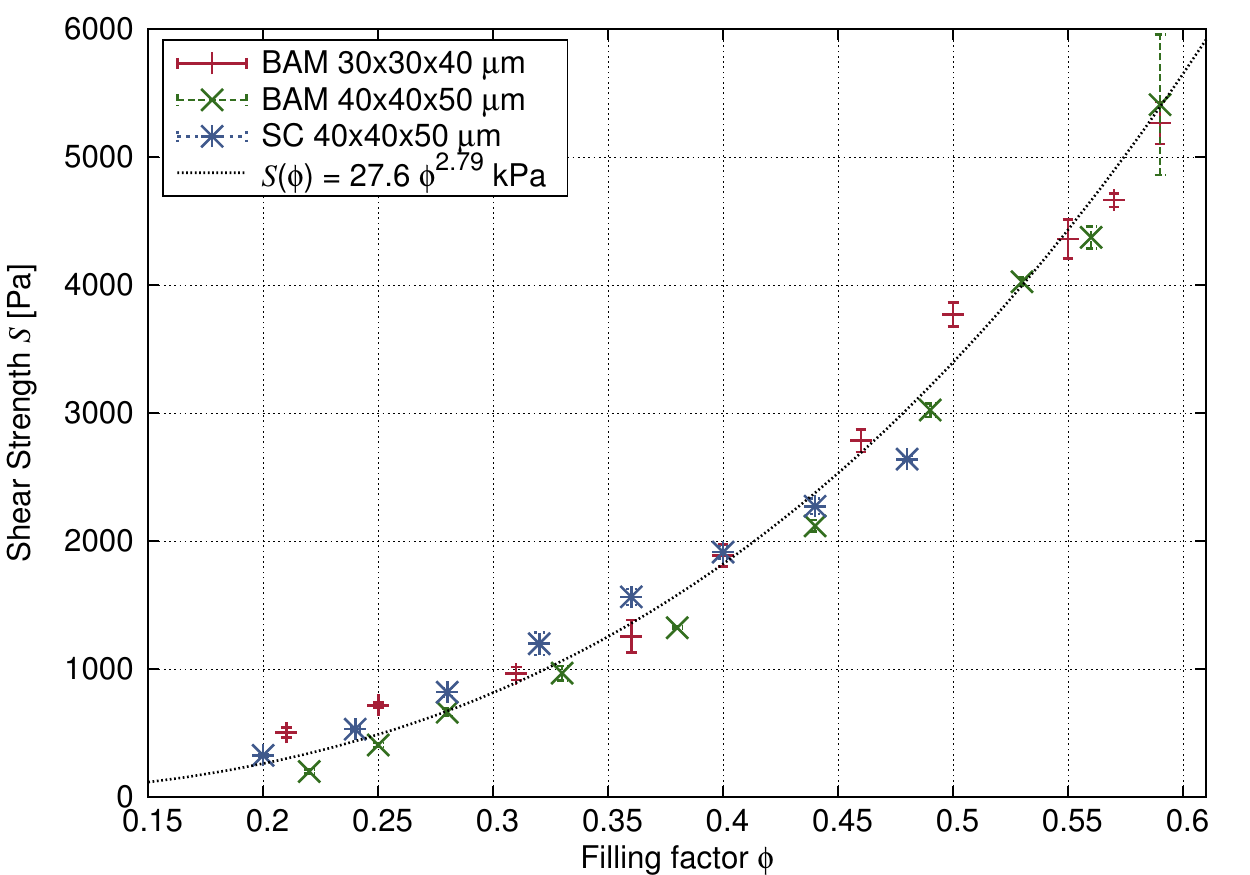}}
\caption{Comparison of the relation between the filling factor $\phi$ and shear strength of different sample types and sizes. The error bars have been determined by averaging the results from six different samples. The black dotted line has been obtained by fitting a power law to the results from BAM and static compaction aggregates with an edge length of $40 \times 40 \times 50\,\mathrm{\mu m}$.}
\label{fig:shear_strength}
\end{figure}

Owing to the computational demand of the simulations the size of our samples is limited to values below $0.1\,\mathrm{mm}$. To study the dependency of our results on the sample size we prepared both BAM and static compaction aggregates with different edge lengths. For each data point six different samples with equal statistical properties have been generated.

Some of the results are shown in Fig.\,\ref{fig:shear_strength}. As we can see, the results of the different sample sizes do not alter significantly. In order to check whether the length of the sample in direction of the shearing motion is sufficient we also performed simulations for sample sizes of $80 \times 40 \times 50\,\mathrm{\mu m}$ and $120 \times 40 \times 50\,\mathrm{\mu m}$. Owing to the huge number of particles these simulations took several weeks. Therefore we restricted the values of the filling factor to $\phi = 0.33$ and $\phi = 0.49$. The deviation to the values obtained from the smaller $40 \times 40 \times 50\,\mathrm{\mu m}$ aggregates was $\approx 8-10\%$ for $\phi = 0.33$ and $\approx 2-3\%$ for $\phi = 0.49$. Thus, we may draw the conclusion that the samples are in fact sufficiently large. With the exception of the most compact samples ($\phi = 0.59$) the error bars obtained by averaging the results from the six samples are very small.

Interestingly, we do not observe a significant difference between the static compaction and BAM aggregates as in the case of the tensile strength. As explained in Sect.\,\ref{sec:tensile_strength_results}, owing to their generation process the internal structure of the static compaction aggregates is more favorable to counteract external loading/tension. However, this does not apply to shearing motion that results in different kind of deformation compared to the tensile strength test. Therefore, the two types of samples exhibit similar values for the shear strength.

To provide SPH simulations with an easy to implement model for the shear strength we describe the dependency of shear strength $S$ on the filling factor $\phi$ with a power law $S(\phi) = a \phi^b$. Using the results from BAM and static compaction aggregates of $40 \times 40 \times 50\,\mathrm{\mu m}$ edge length we obtained (see Fig.\,\ref{fig:shear_strength})
\begin{eqnarray}S(\phi) = 21.7 \phi^{2.65}\,\mathrm{kPa}.\label{eq:ss_fit}\end{eqnarray}

\subsection{Comparison with the SPH continuum model}

One objective of the present investigations has been the comparison of the resulting strengths with those adopted in the SPH simulations by \citet{2010A&A...513A..58G}. To model shear failure, \citet{2004Icar..167..431S} introduced a von Mises yielding criterion in his SPH simulations. The required shear strength is in principle equivalent to the shear strength obtained in our calculations. \citet{2009ApJ...701..130G} calibrated their SPH model in an extensive process comparing simulation results with laboratory experiments. They found a new representation for the dynamic compressive strength,
\begin{equation}
\Sigma_\mathrm{SPH}(\phi) = 13\left(\frac{\phi_2 - \phi_1}{\phi_2-\phi}-1\right)^{0.58\cdot\ln 10}\,\mathrm{kPa}
\end{equation}
with $\phi_1 = 0.12$ and $\phi_2 = 0.58$,
and they chose the tensile strength according to \citet{PhysRevLett.93.115503},
\begin{equation}
T_\mathrm{SPH}(\phi) = 10^{2.8+1.48\,\phi}\,\mathrm{Pa}.
\label{eq:sphts_fit}
\end{equation}
For the shear strength, no experimental data have been available. Therefore the shear strength was adopted following \citet{2004Icar..167..431S} according to
\begin{equation}
S_\mathrm{SPH} = \sqrt{\Sigma_\mathrm{SPH}T_\mathrm{SPH}}.
\label{eq:sphss_fit}
\end{equation}

\begin{figure}
\resizebox{\hsize}{!}{\includegraphics{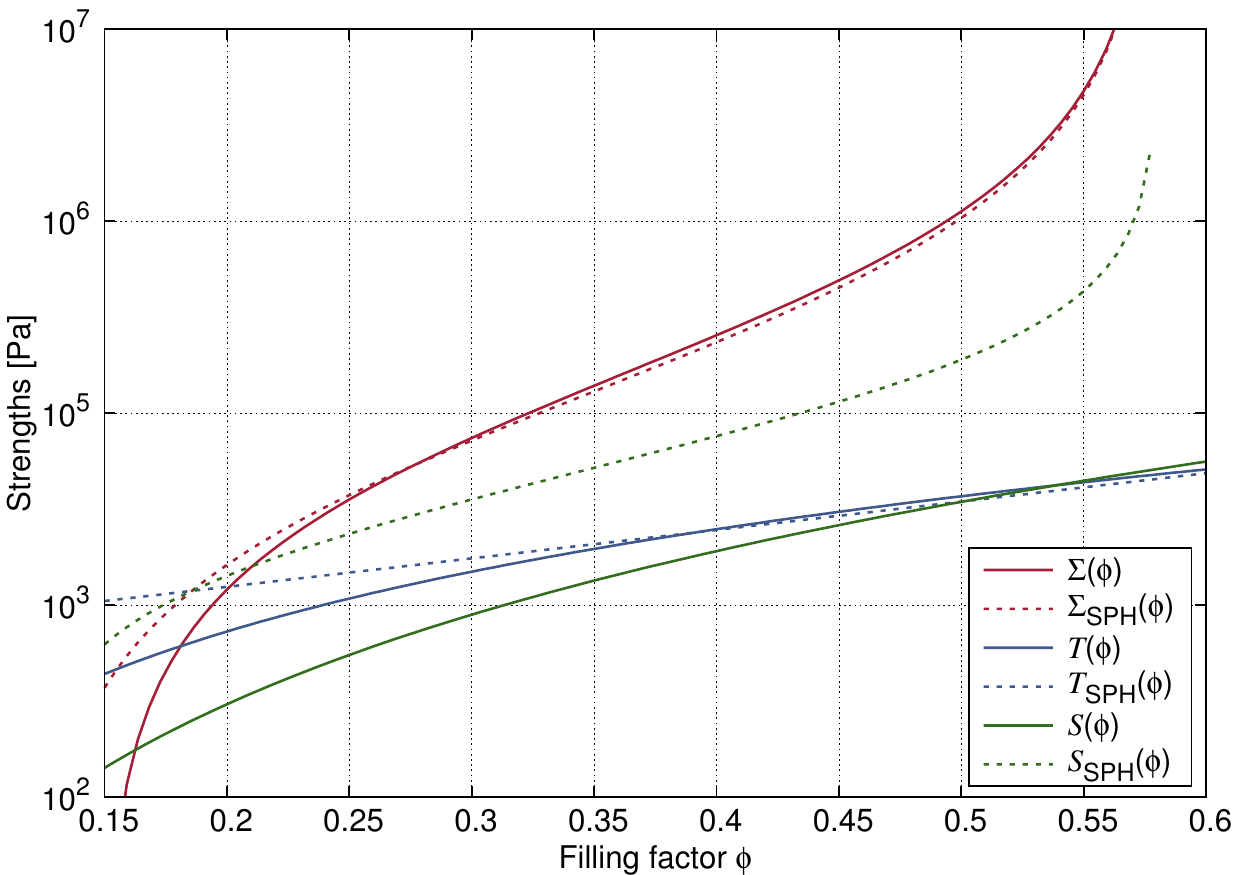}}
\caption{ Comparison of the fit curves for the compressive $\Sigma$, tensile $T$, and shear strength $S$ derived in this work, and the corresponding functions $\Sigma_\mathrm{SPH}$, $T_\mathrm{SPH}$, and $S_\mathrm{SPH}$ adopted in the SPH code by \citet{2010A&A...513A..58G}. The compressive strength $\Sigma(\phi)$ has already been determined in earlier work \citep{2012A&A...541A..59S}. }
\label{fig:sph_strengths}
\end{figure}

In Fig.\,\ref{fig:sph_strengths} the fit curves of tensile strength $T(\phi)$, eq.~(\ref{eq:ts_fit}), and shear strength $S(\phi)$, Eq.~(\ref{eq:ss_fit}), are compared to the corresponding values of the SPH model, $T_\mathrm{SPH}(\phi)$, Eq.~(\ref{eq:sphts_fit}), and $S_\mathrm{SPH}(\phi)$, eq.~(\ref{eq:sphss_fit}). As can be seen, the tensile strength curves match rather well. This emphasizes that the present molecular dynamics method is well suited to model highly porous aggregates. The shear strength curves, however, differ by nearly one order of magnitude. This indicates that the approach of Sirono (Eq.~\ref{eq:sphss_fit}) for the SPH shear strength, which is not based directly on laboratory experiments, might be inappropriate. But during the calibration process it was found already that the SPH simulation results for the chosen reference problems only depend weakly on the exact values of the shear strength \citep{2009ApJ...701..130G}.

\section{Conclusions}

This work supports the observation of \citet{2013A&A...551A..65S} that the sample generation method influences its mechanical behavior significantly. Whereas the bouncing behavior of microscopic dust aggregates differs little for BAM and static compaction aggregates they do behave differently when external strain is applied (see Fig.\,\ref{fig:tensile_strength_cake_types}). It is important to keep this in mind when comparing numerical simulations to laboratory results.

Observing the transition from ductile to brittle behavior for coordination numbers of $\approx 6$ is very interesting. It certainly influences the outcome of collisions as well. For brittle aggregates fragmentation will play a significantly larger role. 

In this work we determined simple power laws to describe the relation between the tensile strength (see Eq.\,\ref{eq:ts_fit}) or shear strength (see Eq.\,\ref{eq:ss_fit}) and the porosity. In combination with earlier work on the the compressive strength \citep{2012A&A...541A..59S} it provides a complete description when the inelastic regime is entered upon deformation of porous dust aggregates. Since the dissipation of the kinetic impact energy is critical, this knowledge is crucial for continuum simulations of collisions of macroscopic porous aggregates.

Comparing with a special SPH model, we find that our tensile strength agrees well with the tensile strength adopted in the SPH code. The same holds for the compressive strength as found in earlier work \citep{2012A&A...541A..59S}. However, the shear strength differs significantly. Future analysis has to show whether our improved relation for the shear strength will have fundamental impact on the SPH simulation results, or whether the shear strength only alters details in the simulations, as might be indicated by previous work.

\begin{acknowledgements}
A.\,Seizinger acknowledges the support through the German Research Foundation (DFG) grant KL 650/16. The authors acknowledge support through DFG grant KL 650/7. Additional support through the German Research Foundation (DFG) through grant KL 650/11 within the Collaborative Research Group FOR 759: {\it The formation of Planets: The Critical First Growth Phase} is acknowledged.

We thank the anonymous referee for pointing out possible misunderstandings and helping to improve the quality of the paper.
\end{acknowledgements}

\bibliographystyle{aa}
\bibliography{references}

\end{document}